\documentclass[useAMS,usegraphicx]{mn2e}
\newcommand{\apj}{ApJ}
\newcommand{\apjl}{ApJL}

\begin{document}
\title 
{Estimate of the total kinetic power and age of 
extragalactic jet
by its cocoon dynamics: The case of Cygnus A}

\author[]{M. Kino and N. Kawakatu\\
SISSA, via Beirut 2-4, 34014 Trieste, Italy}
\date{submitted to MNRAS ---- --- ---}
\twocolumn

\maketitle

\begin{abstract}

We examine the constraints imposed on the total 
kinetic power ($L_{\rm j}$) 
and the age ($t_{\rm age}$) 
of relativistic jets in FR II radio sources
in a new way. 
We solve the dynamical expansion of its cocoon
embedded in the intra-cluster medium (ICM)
and 
obtain the analytic solution of its physical quantities in terms of  
$L_{\rm j}$ and $t_{\rm age}$.
The estimate of $L_{\rm j}$ and $t_{\rm age}$ is done
by the comparison of the model and the observed 
shape of the cocoon.
The analysis is focused on Cygnus A and we find that 
(i) the source age is 
$3\ {\rm Myr} <t_{\rm age}<30\ {\rm Myr}$ and 
(ii) the total kinetic power of the jet
is estimated as 
$0.2\times 10^{46}{\rm erg \ s^{-1}} 
<L_{\rm j}<1\times 10^{48} {\rm erg \ s^{-1}}$
which is larger than  $1\%$ of the 
Eddington luminosity of Cygnus A. 

\end{abstract}


\section{INTRODUCTION}

After the detections 
of inverse Compton component in $X$-ray  
from FR II radio sources, 
our knowledge of energetics, 
especially on the kinetic power of non-thermal electrons, 
is progressed in recent years 
(e.g., Leahy and Gizani 2001; 
Harris and Krawczynski 2002;
Hardcastle at al. 2002; 
Kataoka et al. 2003; Croston et al. 2005). 
When we explore further physical conditions 
in the jets, we encounter a big difficulty 
to constrain on the fraction 
such as thermal electrons and/or protons 
co-existing with non-thermal electrons
because it is hard to observe these component 
(e.g., Celotti et al. 1998; Sikora and Madejski 2002).
This problem prevented us from estimating
the {\it total} mass and energy flux ejected from a
central engine.
To conquer this, a simple procedure is proposed 
in the study of jets in FR II sources in
Kino and Takahara (2004) (hereafter KT04).
In the pressure and mass density of jet,
the contributions from the invisible components 
are also involved. Hence the 
rest mass density and energy density estimated from the
shock dynamics definitely prove the quantities
of total plasma.
Also for radio bubbles in cluster cores,
the similar dynamical approach 
has been adopted to constrain on their physical state
(Fabian et al. 2002; Dunn and Fabian 2004).

In this paper, we will explore 
the cocoon dynamics in FR II radio sources 
as a robust tool to know 
the total kinetic power $L_{\rm j}$ 
and source age $t_{\rm age}$.
In \S2, we discuss our analytical solution of 
cocoon expansion which includes the effect of radial dependence of
the rest mass density of the surrounding Intra Cluster Medium (ICM).
In \S3, by comparing the analytic solution
and observation of the cocoon,
we explore the kinetic power of Cygnus A, which
is one of the best-studied FR II source.
Conclusions are given in \S4.

\section{Dynamics of Cocoon}

We deal with the cocoon dynamics in FR II radio source.
The adopted basic equations are almost the 
same as those in Begelman and Cioffi (1989)
(hereafter BC89).
The main differences between BC89 and the present work are 
(i) we explicitly solve the physical quantities as a function
of $L_{\rm j}$, and $t_{\rm age}$,
(ii) we take account of the effect of
radial profile of the mass density of ICM, and, 
as a result,
(iii) the growth of the cocoon head  $A_{\rm h}$ 
is consistently solved from the basic equations.

\subsection{Basic Assumptions}

Our main assumptions are as follows;
(1) We limit our attention a jet relativistic speed ($v_{j}=c$),
(2) The jet supplies a constant $L_{\rm j}$ in time, 
(3) We focus on the over-pressured cocoon phase, and
(4) We assume that the  magnetic fields are passive
and ignore their dynamical effect.
For (1), although it is still under debate, 
some jets are suggested to be relativistic 
(e.g., Tavecchio et al. 2000; Celotti, Ghisellini, \& Chiaberge 2001).
Since little is known about 
the evolution of $L_{\rm j}$,
the assumption (2) is adopted as a first-step working hypothesis.
The assumption of (3) is automatically guaranteed by the 
sideways expansion of the cocoon (e.g., Cioffi and Blondin 1992). 
The assumption of (4) is based on
the results that a multi-frequency analysis of radio galaxies 
show that the energy density of magnetic field tend to be smaller
than that of non-thermal electrons 
(e.g., Leahy and Giani 2001; Isobe et al. 2002).

\subsection{Basic Equations}

Model parameters are
$L_{\rm j}$ and $t_{\rm age}$.
Unknown physical quantities are
$v_{\rm h}$, 
$v_{\rm c}$,
$P_{\rm c}$, and
$A_{\rm h}$ (or $A_{\rm c}$)
which are
the advance velocity of the cocoon head,
the velocity of cocoon sideways expansion,
the pressure of the cocoon,
the cross sectional area of the head part of the cocoon
(or the cross sectional area of cocoon body),
respectively (see Fig. 1).
Equation of motion along the jet axis,
and sideways expansion, and
energy conservation in the cocoon 
are given by
\begin{eqnarray}
\frac{L_{\rm j}}{v_{\rm j}}=
\rho_{\rm a}(r_{\rm h})v_{\rm h}^{2}(t)A_{\rm h}(t),
\end{eqnarray}
\begin{eqnarray}
P_{\rm c}(t)=
\rho_{\rm a}(r_{\rm c}) \
v_{\rm c}(t)^{2}  ,
\end{eqnarray}
\begin{eqnarray}\label{eq:energyeq}
\frac{P_{\rm c}(t)V_{\rm c}(t)}{\hat{\gamma_{\rm c}}-1}
\simeq 2 L_{\rm j}t    ,
\end{eqnarray}
where
$r_{\rm h}(t)=\int_{t_{\rm min}}^{t} v_{\rm h}(t')dt'$, 
$r_{\rm c}(t)=\int_{t_{\rm min}}^{t} v_{\rm c}(t')dt'$, 
$V_{\rm c}(t)=2\int_{t_{\rm min}}^{t} 
A_{\rm c}(t')v_{\rm h}(t')dt'$, 
$t_{\rm min}$, and 
$\hat{\gamma}_{\rm c}$ are
the length from the center of 
the galaxy to the head of the cocoon, 
the radius of the cocoon body, 
the volume of the cocoon,
the start time of source evolution and 
specific heat ratio of the plasma inside the cocoon,
respectively. 
The declining  mass density 
of ICM $\rho_{\rm a}$  is given by
$\rho_{\rm a}(r)
={\bar\rho}_{\rm a}(r/r_{0})^{-\alpha}$
where $r_{0}$ and 
$\bar{\rho}_{\rm a}$ are
reference position and
the ICM mass density at $r_{0}$, respectively.
We set $r_{0}$ as $r_{\rm h}(t_{\rm age})$ in throughout this paper.
Most of the kinetic energy is deposited in the cocoon,
which is  initially suggested by Scheuer (1974) and 
recent studies of hot spots also shows the 
radiative efficiency is very small (e.g., KT04).
Following to Cioffi and Blondin (1992),
we add the factor of $1/({\hat \gamma}_{c}-1)$ 
in Eq. (\ref{eq:energyeq}) to express the amount of 
the deposited internal energy.
In other words, this corresponds to
the neglect of the $PdV$ work because of its smallness.

The numbers of quantities are $4$, while 
those of basic Eqs. are $3$.
Hence, we set $A_{\rm c}(t) \propto t^{X}$ 
where $X$ as a free parameter.
We can constrain on the value of $X$
from observations. A specific case is shown in \S 3.
Hence, we obtain 
$v_{\rm h}$, 
$v_{\rm c}$,
$P_{\rm c}$, and
$A_{\rm h}$ by using a free parameter $X$.

As a subsidiary equation,
the area of the cocoon body is given by
$A_{\rm c}(t)=
\pi\left(
\int_{t_{\rm min}}^{t} v_{\rm c}(t')dt'
\right)^{2}$.
It is clear that the change of the unknown
from $A_{\rm c}$ to $A_{\rm h}$ does not change the result. 
However note that 
when we choose $A_{\rm h}$ as an unknown instead,
we cannot obtain the solution for $\alpha=2$.

\subsection{Analytic Solution}

We assume that physical quantities have 
a power law dependence in time and the 
coefficient of each physical quantity is barred quantity.
Each quantity has the form of 
$A={\bar A} ~ (t/t_{\rm age})^{Y}$ where $Y$ 
is an arbitrary index.
The time evolution of $v_{\rm c}$ is 
\begin{eqnarray}
v_{\rm c}(t)=
{\bar v}_{\rm c}
\left(\frac{t}{t_{\rm age}}\right)^{0.5X-1}
=
\frac{{\bar A}_{\rm c}^{1/2}}{t_{\rm age}}
{\cal C}_{\rm vc}
\left(\frac{t}{t_{\rm age}}\right)^{0.5X-1},
\end{eqnarray}
for a given  $A_{\rm c}$.
With this, the analytic form of cocoon quantities 
in decreasing ICM density is obtained as follows;
\begin{eqnarray}
P_{\rm c}(t)=
{\bar\rho}_{\rm a} {\bar v}_{\rm c}^{2}
{\cal C}_{\rm pc}
\left(\frac{{\bar v}_{\rm c}}{v_{0}}\right)^{-\alpha}
\left(\frac{t}{t_{\rm age}}\right)^{X(1-\alpha/2)-2},
\end{eqnarray}
\begin{eqnarray}
v_{\rm h}(L_{\rm j},t)=
\frac
{L_{j}}
{{\bar\rho}_{\rm a}{\bar v}_{c}^{2}{\bar A}_{\rm c}}
{\cal C}_{\rm vh} 
\left(\frac{{\bar v}_{\rm c}}{v_{0}}\right)^{\alpha}
\left(\frac{t}{t_{\rm age}}\right)^{X(-2+0.5\alpha)+2},
\end{eqnarray}
\begin{eqnarray}\label{eq:ah}
A_{\rm h}(L_{j},t)=
\frac{L_{\rm j}}
{v_{\rm j}{\bar\rho}_{\rm a} 
{\bar v}_{\rm h}^{2}}
{\cal C}_{\rm ah}
\left(\frac{{\bar v}_{\rm h}}{v_{0}}\right)^{\alpha}
\left(\frac{t}{t_{\rm age}}\right)
^{X(\alpha-2)(-2+0.5\alpha)+3\alpha-4} ,
\end{eqnarray}
where 
${\cal C}_{\rm vh}=
(\hat{\gamma}_{c}-1)[3-(1-0.5\alpha)X](0.5X)^{-\alpha}$,
${\cal C}_{\rm vc}=0.5X/ \pi^{1/2}$,
${\cal C}_{\rm pc}=(0.5X)^{\alpha}$, and 
${\cal C}_{\rm ah}=[X(-2+0.5\alpha)+3]^{-\alpha}$, respectively.
$v_{0}\equiv r_{\rm h}/t_{\rm age}$ corresponds to
the head speed for constant velocity in time. 

Here we use 
the conditions of $0.5X>0$ and $X(-2+0.5\alpha)+3>0$ 
which make the contribution at  $t_{\rm min}$ 
in the  integrations of  $r_{\rm h}$ and  $r_{\rm c}$
small enough.
The case we focus on in \S 3
is that $X=12/7$ and $\alpha=1.5$, which 
clearly satisfies these conditions.

First, let us consider the evolution of cocoon. 
The growth of both 
$A_{\rm h}$ and $A_{\rm c}$ must be positive.
As for the cocoon expansion speeds,
three different behaviors are theoretically
predicted such as
(I) {\it accelerated-head} ($X(-2+0.5\alpha)+2>0$),
(II) {\it constant-head} ($X(-2+0.5\alpha)+2=0$), and
(III) {\it decelerated-head} ($X(-2+0.5\alpha)+2<0$).
The case of (I), (II), and (III) correspond to
$X<1$, $X=1$, and $X>1$ for $\alpha=0$,   
while in the case of $\alpha=2$ (I), (II), and (III) correspond to
$X<2$, $X=2$, and $X>2$, respectively. 
Related to this,
it is useful to express the aspect ratio of the cocoon  
$\frac{r_{\rm c}}{r_{\rm h}}\equiv {\cal R}$ as a function
of time. This is written as 
\begin{eqnarray}\label{eq:ratio}
{\cal R}(t)=
\frac{X(-2+0.5\alpha)+3}{0.5X}
\frac{{\bar v}_{\rm c}}{{\bar v}_{\rm h}}
\left(\frac{t}{t_{\rm age}}
\right)^{X(2.5-0.5\alpha)-3}  .
\end{eqnarray}
%
It is worth to note that
the solution describes 
not only the self-similar evolution 
(e.g.,  Kaiser and Alexander 1997;
Bicknell, Dopita and O'Dea 1997)
but also the non self-similar one. 
Although it is not concerned with in this paper, 
the evolution of ${\cal R}(t)$ may probe 
the evolution of radio galaxies such  as 
``compact symmetric objects (CSO)''  
which are thought to  be the progenitors of the FR II source 
(e.g., Fanti et al. 1996; Readhead et al. 1996a).
Note that, observationally, the large deviation from 
${\cal R}\sim{\cal O}(1)$ does not seem to be unnatural
for actual radio sources (e.g., Readhead et al. 1996b).

Next, let us consider
the $\alpha$ and $X$ dependences on the quantities
based on Eqs. (1), (2) and (3).
Here we fix the physical quantities at $t=t_{\rm age}$.
With regard to the $\alpha$ dependence in fixed $X$,
larger $\alpha$ leads to
larger $\rho_{\rm a}$,
slower $v_{\rm h}$,
same $v_{\rm c}$,
larger $P_{\rm c}$, 
during $t<t_{\rm age}$.
We can understand it as follows.
Larger $\alpha$ 
leads to stronger the deceleration effect 
on the head speed 
$v_{\rm h}$ due to larger 
$\rho_{\rm a}$. 
Larger $\alpha$ also predict larger
$P_{c}$ in order to keep the same velocity of 
sideways expansion $v_{\rm c}$.
In Fig. \ref{fig:tla}, we show the effect of varying 
$\alpha$ when fixing $X=1$. 
It shows that  larger $\alpha$ requires
larger $L_{\rm j}$ in order to plow the larger amount of  
ICM. 

Next we consider the $X$ dependence.
In fixed $\alpha$, larger $X$ lead to  
faster $v_{\rm h}$,
slower $v_{\rm c}$,
smaller $P_{\rm c}$, and
smaller $A_{\rm h}$
during $t<t_{\rm age}$.
We can explain it as follows.
Lager $X$ 
leads to slower sideways expansion
$v_{\rm c}$ and smaller 
$A_{\rm c}$ by definition.
From the equation of motion to the sideways expansion,
it is clear that smaller $P_{\rm c}$ is required. 
To satisfy the energy equation at the same time,
the faster velocity of $v_{\rm h}$ is needed.
This is realized by smaller area head of the cocoon $A_{\rm h}$.

Observationally, little is argued about 
the emission from the cocoon itself.
Recently, Readhead et al. (1996b) 
report the cocoon emission from the CSO 2352+495.
It is visible in the 610 MHz image and 
the aspect-ratio is about ${\cal R}(t_{\rm min})\sim 1/2$.
Because of the lack of our knowledge 
of the cocoon emissions,
here we examine the case of 
${\cal R}(t_{\rm min})=$1, 1/2, and 1/4
for the moment.
In Fig. \ref{fig:tlx},
we show the effect of 
varying $X$ with fixing 
$\alpha=1.5$ and
${\cal R}(t_{\rm age})=1/2$.
The case of $X=$ 1.62, 12/7, and 1.80
correspond to the 
${\cal R}(t_{\rm min})=$1, 1/2, and 1/4
respectively. We can verify the shift of 
$L_{\rm j}$ to smaller range as the $X$ increases.

Finally, we compare our solution with the previous works.
The solution for  flat ICM density by BC89 model 
corresponds to the case of  $X=1$ and $\alpha=0$
and the quantities are written as  
$v_{\rm c}\propto t^{-1/2}$,
$v_{\rm h}\propto const.$,
$P_{\rm c}\propto t^{-1}$, 
$A_{\rm h}\propto const$.
The absolute values of each physical quantities
are also agree with those in BC89 when
omitting $1/({\hat \gamma}_{c}-1)$ and replace 
${\cal C}_{\rm vc}=1/\pi^{0.5}$,
${\cal C}_{\rm vh}=1$,
${\cal C}_{\rm pc}=1$, and
${\cal C}_{\rm ah}=1$.
It is included in Fig. \ref{fig:tla} the case of BC89
with the parameters of 
$\alpha=0$, 
$X=1$, and 
$A_{\rm h}=30$ kpc$^{2}$. 
As explained before, the larger (smaller) $\alpha$ predict 
smaller (larger) $L_{\rm j}$ because of the dilute (dense) ICM. 
The comparison with the analytic model
and numerical studies, 
is also important issue.
By Cioffi and Blondin (1992), 
the issues of  ``head cross section growth'' and ``decreasing
head velocity''  is assessed by
hydrodynamic simulation and
their result shows $A_{\rm h}\propto t^{0.4}$.
As a good example, our solution of $X\simeq 1.1$ describes 
the studies Nath (1995) corresponds to
case of flat solution above
$v_{\rm h}\propto t^{-0.2}$,
$v_{\rm c}\propto t^{-0.45}$,
$P_{\rm c}\propto t^{-0.9}$, and
$A_{\rm h}\propto t^{0.4}$.
Concerning $A_{\rm c}$,
since most of numerical studies have mainly focused
on the propagation of cylindrical geometry jets 
(e.g., Marti et al. 1997; Clarke, Harris and Carilli 1997),
it remains as an important future work
on the time evolution of $A_{\rm c}$.


\section{Total Kinetic Power and Source Age of Cygnus A}

Here
we explore $L_{\rm j}$ and $t_{\rm age}$ by matching
the observed cross-sectional areas and lengths of a cocoon 
(i.e.,  
$r_{\rm h}$, 
$r_{\rm c}$, 
$A_{\rm h}$, and 
$A_{\rm c}$).
In this paper,
we focus on  the archetypal radio galaxy Cygnus A
(Carilli and Barthel 1996; Carilli and Harris 1996 for reviews).

First, we estimate the typical values of physical quantities.
Concerning the mass density profile of ICM,
we adopt $\alpha=1.5$ based on 
Reynolds and Fabian (1996) and Smith et al. (2002).
We examine the case of $X=12/7$ as a fiducial one
which predict the constant ${\cal R}$ in time.
Other observed quantities 
${\bar \rho}_{\rm a}=0.5\times 10^{-2} m_{p} \ \rm g \ cm^{-3}$, and
$r_{\rm h}=60\ \rm kpc$, 
based on 
Carilli et al. (1998), 
Arnaud et al.(1984), and  
Smith et al. (2002).
Here we assume $\hat{\gamma}_{c}=4/3$.
Using these quantities, we obtain 
\begin{eqnarray}
\beta_{\rm h}(t)=
8.36 \times 10^{-3}\ 
L_{\rm j,46}
t_{20}^{2}
\left(\frac{t}{t_{20}}\right)^{-1/7},
\end{eqnarray}
\begin{eqnarray}
\beta_{\rm c}(t)=
6.84\times 10^{-3}\ 
t_{20}^{-1}
\left(\frac{t}{t_{20}}\right)^{-1/7},
\end{eqnarray}
\begin{eqnarray}
P_{\rm c}(t)=
4.78 \times 10^{-10}
t_{20}^{-2}
\left(\frac{t}{t_{20}}\right)^{-11/7}
\ \rm dyne \ cm^{-2} \, ,
\end{eqnarray}
\begin{eqnarray}
A_{\rm h}(t)= 66.4 \ 
L_{\rm j,46}^{1/2}
t_{20}^{1/2}
\left(\frac{t}{t_{20}}\right)^{11/7}
\ \rm kpc^2,
\end{eqnarray}
where 
$\beta_{\rm h}=v_{\rm h}/c$,
$\beta_{\rm c}=v_{\rm c}/c$,
$t_{20}=t_{\rm age}/20\rm Myr$,
$L_{\rm j,46}=L_{\rm j}/10^{46}\rm erg ~s^{-1}$,
with the resultant value 
of ${\cal R}(t_{\rm age})=0.815$ for Cygnus A.
Note that
${\cal R}\propto{\bar \beta}_{\rm h}/{\bar \beta}_{\rm c}$
has a
strong dependence on the age such as
$\propto t_{\rm age}^{3}$,
the allowed source age is fairly restricted.

Next, to clarify the allowed range of 
$t_{\rm age}$ and $L_{\rm j}$, we impose
the following conditions;
the condition which should be satisfied is that
(I) ${\cal R}(t_{\rm age})\sim 0.5-0.7$, from the {\it Chandra} image (Wilson et al. 2000),
(II) cocoon pressure is over-pressured 
$ P_{\rm c}>P_{\rm a}=8\times 10^{-11}\rm dyn~cm^{-2}$
(Arnaud et al. 1984; Smith et al. 2002),
(III) the area size of 
$A_{\rm h}$ lies in the range of 
$ 30 \ {\rm kpc^{2}}< A_{\rm h} < 150 \ {\rm kpc^{2}}$.
The minimum value corresponds to the one adopted in BC89.
From the radio image of Perley, Dreher and Cowan (1984),
we employ the maximum value as $A_{\rm h}=150\ \rm kpc^{2}$,  
which corresponds to the cross sectional area
of the radio lobe at the location of the hot spot. 
In Fig. \ref{fig:tl}, we show the resultant source 
age and total kinetic power of the jet.
The region of the source age larger than $\sim30 \rm Myr$ 
is not allowed by the condition (II).
Larger (smaller) $A_{\rm h}$ predict 
larger (smaller) $L_{\rm j}$ and younger $t_{\rm age}$. 
Obtained values are
$2\times 10^{45} {\rm erg\ s^{-1}} 
<L_{\rm j}<1\times 10^{48}{\rm erg\ s^{-1}}$ and 
$3\ {\rm Myr} <t_{\rm age}<30\ {\rm Myr}$.

Let us consider how the uncertainties
of $L_{\rm j}$ and $t_{\rm age}$ are determined.
From Eqs. (\ref{eq:ah}) and (\ref{eq:ratio}),
$A_{\rm h}
\propto
L_{\rm j}
t_{\rm age}^{2}$ and 
${\cal R}\propto
L_{\rm j}^{-1/(\alpha-4)}
t_{\rm age}^{-3/(\alpha-4)}$ are obtained.
Since $A_{\rm h}$ and ${\cal R}$ have
uncertainties with the factors of 5 and 1.4 respectively, 
the allowed region is mainly controlled by $A_{\rm h}$. 
In the present work, 
in view of the comparison of BC89,
we took the minimum value as $A_{\rm h}=30$kpc$^{2}$.
However, from the hydrodynamical point of view,
it seems natural to suppose that
$A_{\rm h}$ to be the close value to the 
cross sectional area of the radio lobe 
at the distance of head, i.e., $A_{\rm h}=150$ kpc$^{2}$.

\subsection{Comparison with  previous works}

The resultant age well agree with the independent result
of synchrotron age model by Carilli et al. (1991), which claims that
$6\ {\rm Myr} <t_{\rm age}<30 \ {\rm Myr}$. 
The velocity of the hot spot 
$\beta_{\rm hs}\sim 0.06$ corresponds to
the source age of 6 Myr, while  $\beta_{\rm hs}\sim 0.01$ 
corresponds to the source age of 30 Myr.

As a complementary result,
$L_{\rm j}$ estimated in KT04  
in the range of $6\ {\rm Myr} <t_{\rm age}<30\ {\rm Myr}$ 
based on the result 
Carilli et al. (1991) is shown 
in Fig. \ref{fig:tl} (the solid line).
KT04 estimate 
the total kinetic power of the relativistic jet as 
\begin{eqnarray}
L_{\rm j}&=& A_{\rm j}c\Gamma_{\rm j}^{2}
\beta_{\rm j}\rho_{\rm j}c^{2}  \nonumber \\
&=&
A_{\rm j}c
\left(\frac{r_{60}}{t_{\rm age}}
\right)^{2}\rho_{\rm a}(r_{60})\propto A_{\rm j}
 \nonumber
\end{eqnarray}
in the strong relativistic shock limit,
where 
$A_{\rm j}=\pi R_{\rm hs}^{2}$,
$\Gamma_{\rm j}$, 
$\beta_{\rm j}c$,  
$\rho_{\rm j}$, and 
$R_{\rm hs}=2\ \rm kpc$ (Wilson et al. 2000)
are 
the cross-sectional area of the jet
the Lorentz factor,
the velocity, and 
mass density of the jet, and 
the hot spot radius,
respectively.
It should be stressed that  the cocoon model
can predict both $L_{\rm j}$ and $t_{\rm age}$ at the same time, 
while the synchrotron aging model (Carilli et al. 1991)
and the 1D jet model (KT04)
only determine $t_{\rm age}$ or $L_{\rm j}$.
Although these three models are independent, 
they show a reasonable agreement on the values 
of $L_{\rm j}$ and $t_{\rm age}$  with each other 
at least in order-of-magnitude 
in the case of Cygnus A.
At the same time, it is worth to discuss 
a factor of deviation of  the estimate $L_{\rm j}$
by the cocoon model and 1D jet model.
In the same way as KT04,
1D analysis  between the jet and ICM 
take the plane parallel assumption 
(e.g., Begelman, Blandford \& Rees 1984).
The approximation means that $A_{\rm j}$ equals to
$A_{\rm h}$. 
However, $A_{\rm h}$ is supposed to be larger
than  $A_{\rm j}$. 
Hence, the plane parallel approximation
would cause the underestimate of the $L_{\rm j}$ in KT04.

Rawlings \& Saunders (1991) (hereafter RS91) is 
the pioneering work on the correlation between 
the $L_{\rm j}$
and the luminosity of the narrow line regions, which lies 
in close to the central engine.
Then the comparison with the present work and RS91 is 
intriguing issue.
For the sources where synchrotron spectral aging
are available, it is simply by
\begin{eqnarray}
Q
=\frac{E_{\rm eq}}{t_{\rm age}\eta}, \quad \eta=0.5 \nonumber 
\end{eqnarray}
where 
$Q$,
$\eta$, and  
$E_{\rm eq}$ are 
the kinetic power of jet,
a parameter expressing the fraction of the work done on the ICM,  and 
the equipartition energy
with the electrons and the magnetic field field
make an equal contribution to the total energy density
(e.g.,  Miley (1980)), respectively.
We focus on the sample sources where
$t_{\rm age}$ is independently 
obtained by synchrotron aging method.
One difference between the present work
and RS91 is that
we solved the equations of motion and energy equation
(3 Eqs. in total),
while the RS91 only employ energy equation.
Because of this, 
we can eliminate the free parameter $\eta$.
More important and essential difference is 
that $Q$ in RS91 
is not a total kinetic power but 
it is just a equipartition power
(even though they insist that it as a total power). 
We emphasize the advantage of our
work of dealing with the total kinetic power
whilst RS91 only handles the equipartition power 
of extragalactic jets.

Kaiser (2000) (hereafter K00) independently
addressed this quantity by studying 
some bright radio sources involving Cygnus A.
Compared with RS91,
the common advantage is that both K00 and the present work 
can estimate the total kinetic power by directly dealing with
the hydrodynamics.
The way of comparing the observation 
with the model by K00 is different from ours.
K00 matched the surface brightness distribution of 
the cocoon along the jet axis.
The main difference between the present work
and K00 is the derived cocoon pressure during
the matching of the observations and the models.
In the example of Cygnus A, compared with
our estimate of 
$P_{c}\sim 5\times 10^{-10}$ erg cm$^{-3}$,
they tend to estimate smaller $P_{c}$
of order of 
$P_{c}\sim 10^{-12}- 10^{-11}$ erg cm$^{-3}$
(Tables 2, 3 and 4 in K00).
This mainly cause the deviation of the 
derived total kinetic power of Cygnus A as 
$L_{\rm j}\sim {\rm (a \ few)} \times 10^{46}$ erg s$^{-1}$,
whilst K00 derive the total kinetic power with order of 
$L_{\rm j}\sim {\rm (a \ few)} \times 10^{45}$ erg s$^{-1}$.

\subsection{Efficiency of accretion power to kinetic power}

In Tadhunter et al. (2003), 
the mass of the super-massive black hole (SMBH) of Cygnus A 
is reported as $2.5\times 10^{9}M_{\odot}$
which leads to the Eddington luminosity as  
$L_{\rm Edd}=3\times 10^{47}$ erg s$^{-1}$.
From this it follows that
$L_{\rm j}/L_{\rm Edd}\sim 0.01-1$.
On the basis of the observational evidence for the 
accretion-flow origin for the jet 
in AGNs (e.g., Marscher et al. 2002),
it is clear that
the value of $L_{\rm j}/L_{\rm Edd}$ directly shows the required
minimum rate of the mass accretion 
onto the SMBH
normalized by the corresponding Eddington mass accretion rate.

Merloni, Heinz and Di Matteo 2003 
 (see also Maccarone, Gallo and Fender 2003)
examine the disk-jet connection 
by studying  the correlation between
the radio ($L_{\rm R}$)  and the X-ray ($L_{\rm X}$)
luminosity and the black hole mass.
With the large samples of  the stellar mass black holes
and SMBHs,
they claim these sources have the
``fundamental plane'' in the 3 dimensional space of these 
quantities. 
On the way,
the quantity of $L_{X}/L_{\rm Edd}$
is brought up 
to probe the activity of the central engine.
For Cygnus A, Young et al. (2002) reports
$L_{X}\simeq 3.7\times 10^{44} {\rm erg \ s^{-1}}$
which leads to $L_{X}/L_{\rm Edd}\sim 10^{-3}$.
Whereas we recognize the usability of 
the quantity of $L_{X}/L_{\rm Edd}$
to probe the engine activity,
we emphasize that the $L_{X}$
largely depends on the accretion flow model and 
the radiation efficiency at X-ray band. 
On the other hand,
the quantity of $L_{\rm j}/L_{\rm Edd}$
addressed in the present work
does not have the model dependence
and directly shows us the engine activity.

Maraschi \& Tavecchio (2003) also
addressed this 
quantity by collecting
large samples of blazars.
They estimated the jet power
from the blazar's non-thermal spectrum energy distribution,
which are well understood as synchrotron plus
inverse Compton 
(e.g., 
Maraschi et al. 1992; 
Sikora et al. 1994; 
Blandford \& Levinson 1995;
Kino et al. 2002).
The upper limit of the accretion power is 
inferred from the 
luminosity of observed broad emission lines.
They conclude that
for flat-spectrum radio quasars (FSRQs)
the the total power of the jet is of the same
order as the accretion power.  
Note that 
their estimate of the accretion power from the emission line
is putative way and the estimated 
total power does not include the contribution of protons
associated with thermal electrons which
may cause some underestimate of the total power of the jet.

We again emphasize that
as we see above, the value of
$L_{\rm j}/L_{\rm Edd}$ for Cygnus A
in the present work is obtained
with fewer assumptions.
Moreover it make us a fairly robust probe
of the central engine activity of AGN jets.
The application of our method to 
a large sample of other AGN jets 
surely bring about new and  important
knowledges. 
This is actually our ongoing project.

\section{Conclusion}

Our main conclusions in the present work are as follows;

(i) 
A new method to estimate the total kinetic power of the jet
$L_{\rm j}$
and source age $t_{\rm age}$ in powerful FR II radio sources
is proposed. For that,
the study of cocoon dynamics by BC89 is revisited 
and physical quantities associated with the cocoon
are analytically solved as functions of $L_{\rm j}$ and $t_{\rm age}$.
The comparison of the analytic solution with 
the observed cocoon shape
lead to the $L_{\rm j}$ and $t_{\rm age}$ in general.

(ii)
The analysis is focused on Cygnus A, 
with  the conditions of $0.5\le {\cal R}\le0.7$ and 
$30\ {\rm kpc^{2}}<A_{\rm h}<150\ {\rm kpc^{2}}$.
The estimated age 
$3\ {\rm Myr} <t_{\rm age}<30\ {\rm Myr}$
shows a good agreement of the independently estimate age 
by synchrotron aging model by Carilli et al. (1991).
The estimated total kinetic power
$0.2\times 10^{46}{\rm erg \ s^{-1}}
<L_{\rm j}<1\times 10^{48}{\rm erg \ s^{-1}}$
has also 
a reasonable agreement with the independent 1D jet model of KT04
with the aid of $6\ {\rm Myr} <t_{\rm age}<30\ {\rm Myr}$.

(iii) For Cygnus A,
we find that  the total kinetic power lies in the
range of 
$L_{\rm j}/L_{\rm Edd}\sim 0.01-1$,
while the $X$-ray luminosity 
$L_{X}\simeq 3.7\times 10^{44} {\rm erg \ s^{-1}}$ (Young et al. 2002)
satisfies $L_{X}/L_{\rm Edd}\sim 10^{-3}$.
The result of 
$L_{\rm j}/L_{\rm Edd}\sim 0.01-1$
indicate the lower limit of the 
mass accretion rate, which gives
the crucial hint 
for resolving the jet formation problem.

\section*{Acknowledgments}
We appreciate the helpful comments of 
the referee to improve the paper.
We thank 
A. C. Fabian, D. E. Harris, and D. Schwartz for valuable comments.
M. K. thank A. Celotti, J. Kataoka, N. Isobe
and A. Mizuta for stimulating discussions.
We acknowledge the Italian MIUR and
INAF financial supports.




\begin{figure} 
\includegraphics[width=8cm]{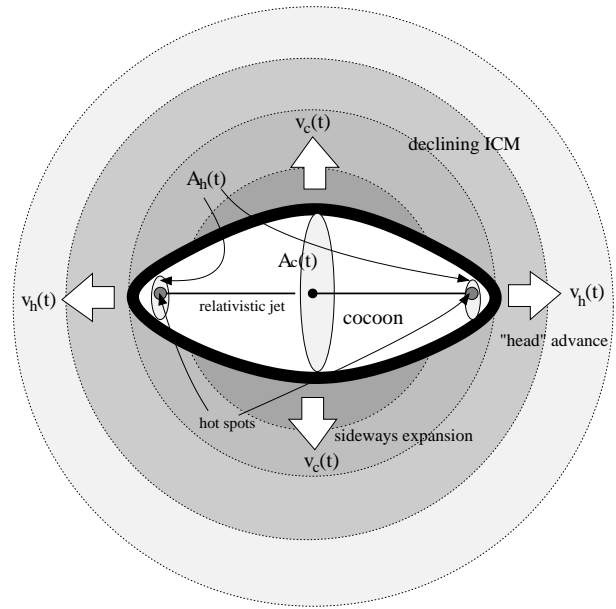}
\caption
{A cartoon of interaction of 
the ICM with declining atmosphere  
and the relativistic jet in FR II radio galaxy. 
As a result, most of the kinetic energy
of jet is deposited in the cocoon and it is 
inflated by its internal energy.}
\label{fig:cocoon}
\end{figure}
\begin{figure} 
\includegraphics[width=8cm]{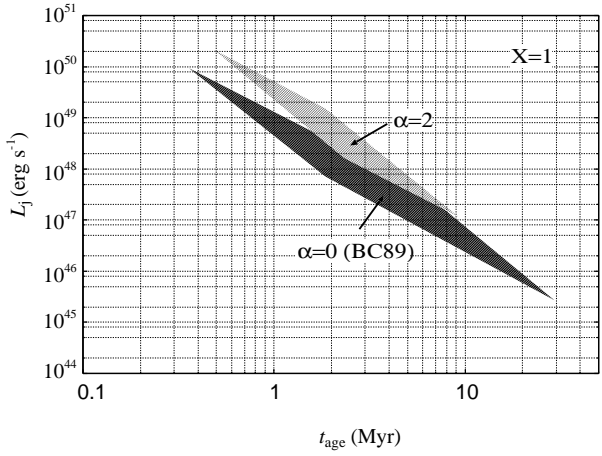}
\caption
{
The 
allowed regions of $L_{\rm j}$ and $t_{\rm age}$
with the different
values of $\alpha$ with $X=1$.
The region is
determined by the constraints of 
$0.5<{\cal R}<1$ and
$30 {\rm kpc^{2}}<A_{\rm h}< 150{\rm kpc^{2}}$.
We examine $\alpha=$0 and 2 here. 
The case of 
$\alpha=0$, and $A_{\rm h}=30 {\rm kpc^{2}}$
by Begelman and Cioffi (1989) is involved in the right-lower part
of the filled region. 
Larger $\alpha$ requires
significantly larger $L_{\rm j}$
for plowing the larger amount of ICM.}
\label{fig:tla}
\end{figure}
\begin{figure} 
\includegraphics[width=8cm]{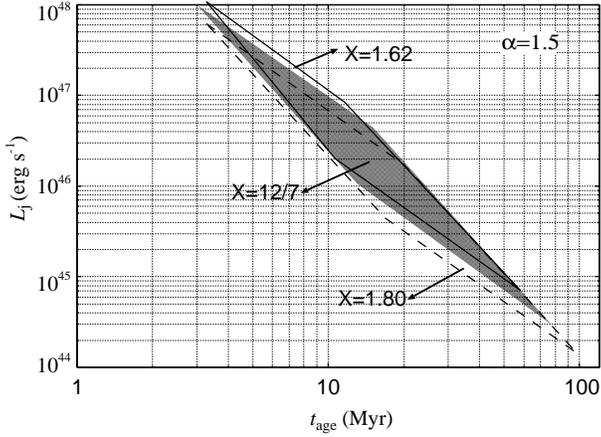}
\caption
{
The allowed regions of $L_{\rm j}$ and $t_{\rm age}$
with the diffferent values of $X$ with $\alpha=1.5$.
Likewise Fig. \ref{fig:tla}, we adopt
$0.5<{\cal R}<1$ and $30 {\rm kpc^{2}}<A_{\rm h}< 150{\rm kpc^{2}}$.
The case of  $X=$1.62, 12/7 and 1.80 are shown here. 
Lager $X$ requires smaller $L_{\rm j}$ corresponding to
the slower velocity of the sideways expand.}
\label{fig:tlx}
\end{figure}
\begin{figure} 
\includegraphics[width=8cm]{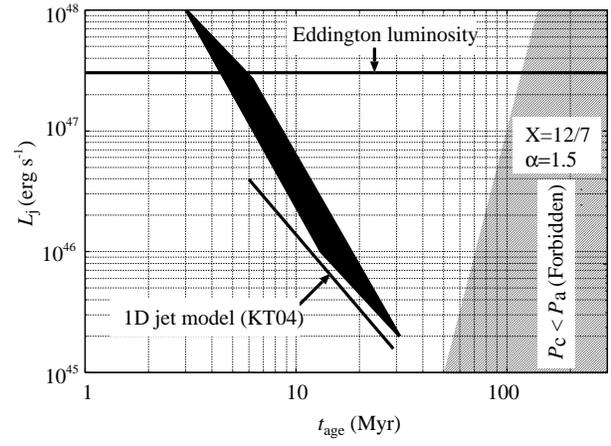}
\caption
{
The allowed region of $L_{\rm j}$ and $t_{\rm age}$
of Cygnus A  (filled in black).
The under-pressured region $P_{\rm c}<P_{\rm a}$ is excluded
by definition. 
The case of $0.5<{\cal R}<0.7$ and
$30 {\rm kpc^{2}}<A_{\rm h}< 150{\rm kpc^{2}}$ is examined. 
As a reference, Eddington luminosity and 
the total kinetic power of jet estimated in KT04
are shown in the thick-solid and solid lines, respectively.}
\label{fig:tl}
\end{figure}

\end{document}